\documentclass[runningheads]{llncs}
\usepackage[T1]{fontenc}
\usepackage[colorlinks,linkcolor=green,anchorcolor=green,citecolor=green]{hyperref}
\usepackage{graphicx}
\usepackage{multirow}
\usepackage{amsfonts,amssymb} 
\usepackage[misc]{ifsym}
\begin{document}
\title{Position-prior Clustering-based Self-attention Module for Knee Cartilage Segmentation}%
%\thanks{Supported by organization x.}}
%
\titlerunning{Position-prior Clustering-based Self-attention Module}
% If the paper title is too long for the running head, you can set
% an abbreviated paper title here
%
\author{Dong Liang\inst{1}, Jun Liu\inst{1}, Kuanquan Wang$^{(\textrm{\Letter})}$\inst{1}, Gongning Luo$^{(\textrm{\Letter})}$\inst{1}, Wei Wang\inst{1}, Shuo Li\inst{2}}%\1{Liang, Dong}, \2{Liu, Jun}, \3{Wang, Kuanquan}, \4{Luo, Gongning}, \5{Wang, Wei}, \6{Li, Shuo}
\authorrunning{D. Liang et al.}
% First names are abbreviated in the running head.
% If there are more than two authors, 'et al.' is used.
%
\institute{Harbin Institute of Technology, Harbin, China\\
\email{\{wangkq, luogongning\}@hit.edu.cn} \\ \and
University of Western Ontario, London, Canada
}
\maketitle              % typeset the header of the contribution
\begin{abstract}
The morphological changes in knee cartilage (especially femoral and tibial cartilages) are closely related to the progression of knee osteoarthritis, which is expressed by magnetic resonance (MR) images and assessed on the cartilage segmentation results. Thus, it is necessary to propose an effective automatic cartilage segmentation model for longitudinal research on osteoarthritis. In this research, to relieve the problem of inaccurate discontinuous segmentation caused by the limited receptive field in convolutional neural networks, we proposed a novel position-prior clustering-based self-attention module (PCAM). In PCAM, long-range dependency between each class center and feature point is captured by self-attention allowing contextual information re-allocated to strengthen the relative features and ensure the continuity of segmentation result. The clutsering-based method is used to estimate class centers, which fosters intra-class consistency and further improves the accuracy of segmentation results. The position-prior excludes the false positives from side-output and makes center estimation more precise. Sufficient experiments are conducted on OAI-ZIB dataset. The experimental results show that the segmentation performance of combination of segmentation network and PCAM obtains an evident improvement compared to original model, which proves the potential application of PCAM in medical segmentation tasks. The source code is publicly available from link: \url{https://github.com/LeongDong/PCAMNet}

\end{abstract}
\section{Introduction}
Osteoarthritis (OA) is a kind of chronic degenerative articular disease that causes disability gradually. Knee cartilage is the soft tissue adhering to the end of the bone surface, whose changes in morphological structure are associated with the progression of OA~\cite{ref_proc1}. Compared to other imaging techniques, magnetic resonance imaging (MRI) shows a higher level of specificity and sensitivity to obtain the biomedical markers of knee cartilage~\cite{ref_article1}. However, manual cartilage segmentation from MRI has demanded more relative knowledge from specialists. The manual segmentation is tedious, time-consuming and brings inter-/intra-observer variations. Thus, there is a demand to design an effective automatic cartilage segmentation method for the longitudinal analysis.

\begin{figure}
\includegraphics[width=\textwidth]{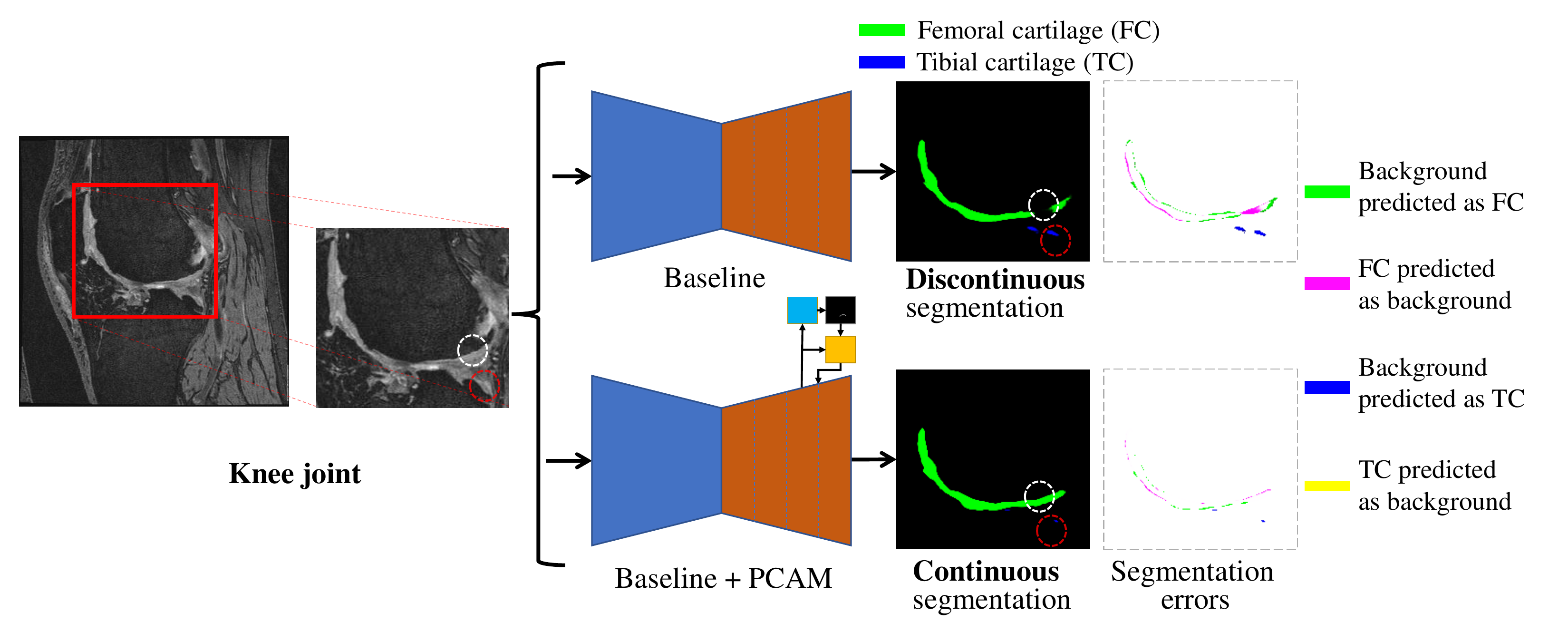}
\centering
\caption{The knee cartilage segmentation results from the primary baseline~\cite{ref_proc4} and the same model~\cite{ref_proc4} modified by PCAM. In white and red circle, the local contextual information of foreground is similar to the background nearby, which causes the segmentation errors and leads to discontinuous segmentation result. The proposed PCAM can relieve the problem and create a more accurate and continuous segmentation result.} \label{fig1}
\end{figure}

With the development of deep learning, convolutional neural network(e.g. U-Net~\cite{ref_proc3}, V-Net~\cite{ref_proc4})has achieved state-of-the-art segmentation results. Although the 3D deep learning model V-Net has exhibited a superior performance on medical image segmentation tasks, directly applying the primary V-Net on knee MR data may generate low accuracy results. As shown in Fig.~\ref{fig1}, the articular structure is complex and the features of tissues around the knee cartilage in MR image are similar to each other, which is difficult to extract continuous knee cartilage accurately from the whole volumetric data. To reduce disturbance brought by the complex background, Ambellan et al.~\cite{ref_article3} proposed a coarse-to-fine scheme with a 2D CNN for coarse segmentation and a 3D CNN for fine segmentation followed by a statistical shape adjustment. Similarly, Gatti et al. ~\cite{ref_article2} adopted 2D U-Net and 3D U-Net in parallel for coarse segmentation and an additional 3D U-Net for fine segmentation. The coarse-to-fine architecture consists of several sub-networks which  brings a huge computational burdens and the input of the behind sub-networks is entirely reliant on the output of the preceding sub-networks. Tan et al. ~\cite{ref_proc2} presented a collaborative multi-agent learning framework in an end-to-end scheme that is still limited to the GPU memory. 
Within 3D MR dataset, the morphological feature of knee cartilage varies greatly. To capture the contextual information of objects with different scales, Zhao et al.~\cite{ref_proc5} presented a pyramid network using multiple dilated convolution kernels. However, this non-adaptive approach ignored the long-range dependencies and did not distinguish the surrounding pixels from different categories leading to discontinuous segmented results. Sinha et al.~\cite{ref_article4} pointed out that the self-attention mechanism exhibits good performance on modeling long-range contextual dependency with a high level of flexibility. Nevertheless, when the feature map is large, this attention module is prone to a heavy computational burden.

To overcome the shortcomings mentioned above, we proposed a novel position-prior and clustering-based self-attention module (PCAM) in CNN for automatic knee cartilage segmentation in MR data. The main contributions of our research are concluded as follows: (a) It is the first time that we applied clustering-based self-attention module on knee cartilage segmentation tasks. And the proposed PCAM can be plugged in network flexibly with less GPU memory consumption and computational burdens. (b) We proposed a position-prior module that excludes false positives on the boundary area of knee cartilage from coarse mask to improve the accuracy of feature clustering in PCAM. (c) The presented PCAM captures long-range contextual information to achieve continuous and accurate knee cartilage segmentation. (d) The segmentation models combining proposed PCAM have obtained performance improvement on the OAI-ZIB dataset.
\section{Method}
\begin{figure}[h]
\includegraphics[width=\textwidth]{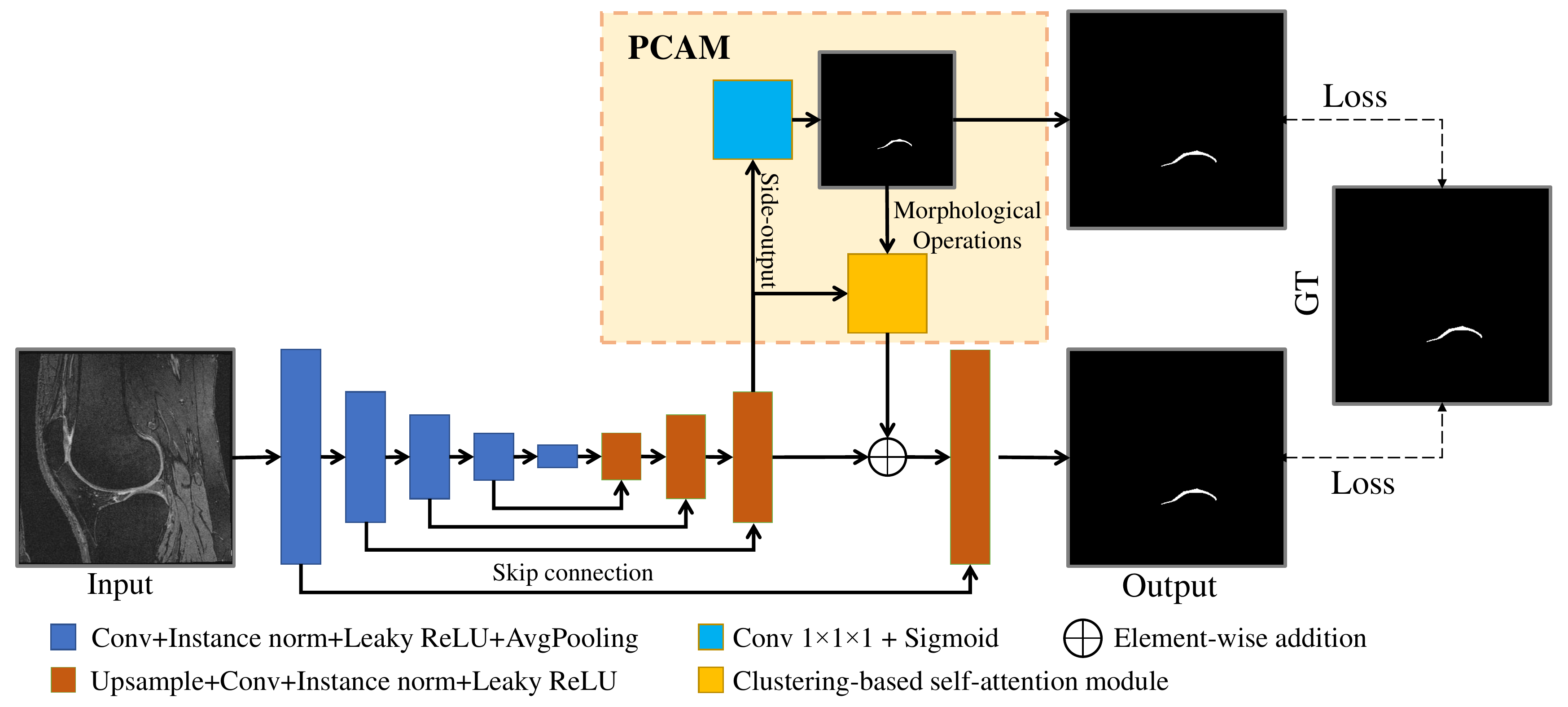}
\caption{The segmentation network architecture with PCAM.} \label{fig2}
\end{figure}

General CNN focuses on the local receptive field while neglecting the long-range dependencies that introduces intra-class inconsistency and discontinuous segmentation~\cite{ref_article4}. The attention module can capture various scale context dependencies and strengthen the relevant feature information to obtain a more accurate result in segmentation tasks. Taking tibial cartilage segmentation as an example, the overview of the network architecture is shown in Fig.~\ref{fig2}. The blue box is the side-output function to produced coarse mask of the knee cartilage that is modified by morphological operations. The feature map together with the modified side-output is then fed back to the clustering-based self-attention module to produce an enhanced feature map.

The PCAM is divided into three parts: position-prior module, clustering-based module and self-attention module, which are illustrated in Fig.~\ref{fig3}. As it is impractical to use true label mask for calculating each class center in corresponding feature map, the output of the segmentation network is applied for class center approximation~\cite{ref_proc6}. However, the knee cartilage only occupies a small area in a large sized MR image, several false positive points in the coarse segmentation result can make the estimated results deviate from the true class centers evidently. Therefore, we adopt modified result with the help of morphological operations to generate precise position prior in this research.
\subsubsection{Position-prior module}
The position-prior module is designed for excluding the false positives of side-output so as to improve the accuracy of feature clustering. Fig.~\ref{fig1} illustrates a common case that V-Net obtains precise segmentation results within the knee cartilage as well as background while fails on the area around the boundary of segmented knee cartilage. In boundary area, the feature distribution is ambiguous because of the low contrast with adjacent tissues and the poor quality of imaging technique. Thus, we divided the predicted probability map into three parts: $M^{boundary}$ (the area around boundary the of predicted knee cartilage); $M^{foreground}$ (the area within predicted knee cartilage); $M^{background}$ (the area except for $M^{boundary}$ and $M^{foreground}$). Areas $M^{foreground}$ and $M^{background}$ are defined as follows:
\begin{equation}\label{e1}
M^{foreground} = \{(x,y,z)|(B)_{(x,y,z)}\subseteq \sigma(F)\}
\end{equation}
\begin{equation}\label{e2}
M^{background} = \{(x,y,z)|(B)_{(x,y,z)}\subseteq 1-\sigma(F)\}
\end{equation}
where $F$ is the feature map and $\sigma(\cdot)$ is the side-output function to generate a predicted probability map; $(x,y,z)$ is the position in $\sigma(F)$; $B$ represents the structure element while function $(B)_{(x,y,z)}$ is the set centered on $(x,y,z)$ containing all elements of $B$. As shown in Fig.~\ref{fig3}, we regard the modified side-output $M^{foreground}$ and $M^{background}$ as position prior, which are then used to assess class centers of foreground and background in feature map, respectively.
\subsubsection{Clustering-based module}
\begin{figure}[h]
\includegraphics[width=\textwidth]{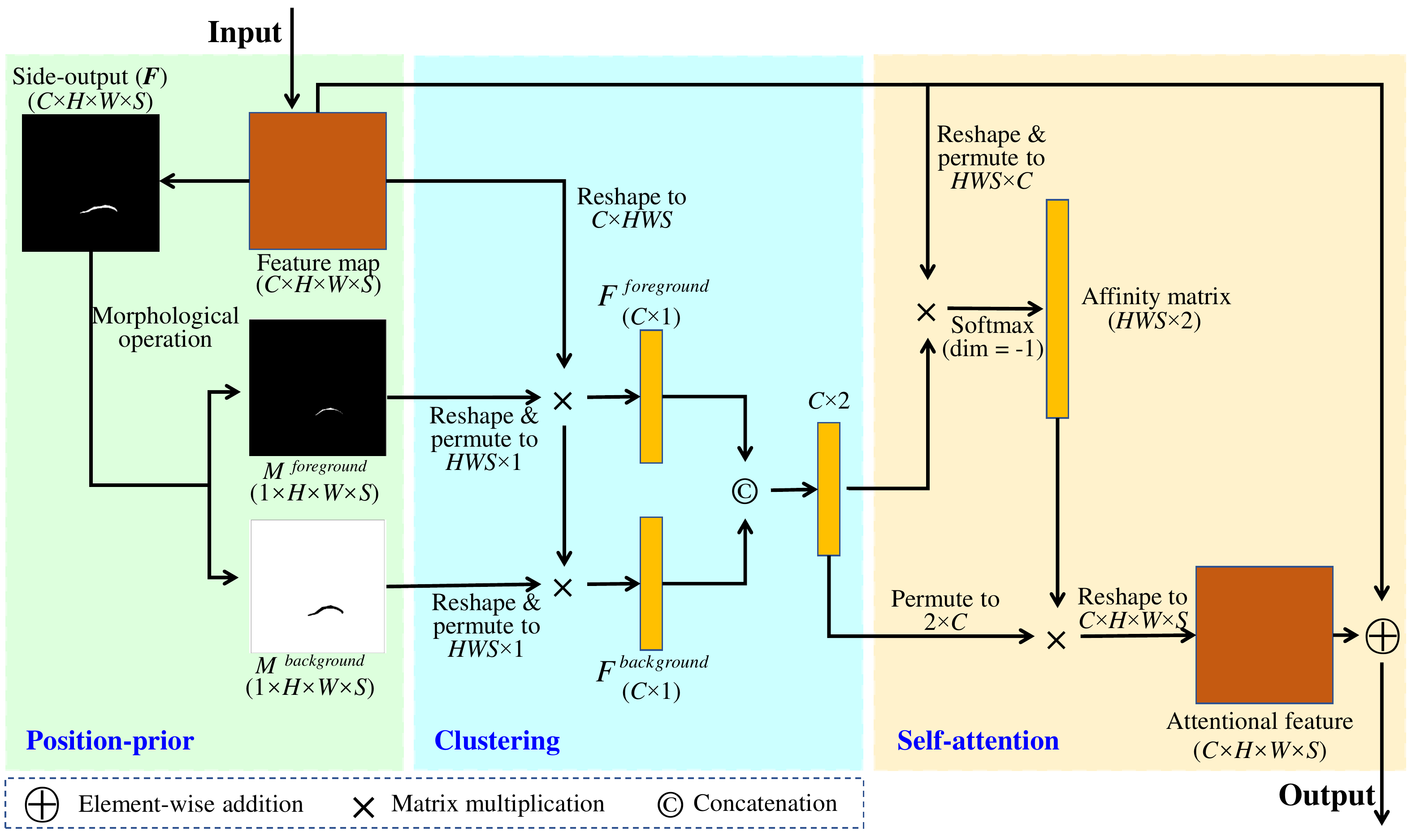}
\caption{The details of position-prior and clustering-based attention module.} \label{fig3}
\end{figure}

The clustering-based module can weaken the influence of segmentation error by averaging all features belonging to the same class in predicted probability map. In addition, the class center contains abundant contextual information as an aggregation of features within each class. Thus, the clustering method is applied in PCAM to compute the similarity between every position in feature map and each class center so as to construct affinity map. The process of clustering is shown in Fig.~\ref{fig3}.
Given the feature map $F\in{\mathbb{R}}^{C\times H\times W\times S}$, foreground mask $M^{foreground}\in{\mathbb{R}}^{1\times H\times W\times S}$ and background mask $M^{background}\in\mathbb{R}^{1\times H \times W\times S}$, $F$ is reshaped to $C\times HWS$ and mask $M^{class}$ is adjusted to shape $HWS\times 1$, where $C, H, W, S$ represent channel, height, width and slice, respectively. The class center of each class is calculated as follows:
\begin{equation}\label{e3}
F^{class} = \frac{\sum_{i=1}^{HWS}M^{class}(i)\cdot F(i)}{\sum_{i=1}^{HWS}M^{class}(i)}
\end{equation}
where the $class$ could be one of the elements from set $\{foreground, background\}$, $F(i)\in \mathbb{R}^{C\times 1}$ is the feature vector in position $i$ of feature map $F$. 
\subsubsection{Self-attention module}
Self-attention module can capture the long-range contextual information to ensure the continuous segmentation result and improve the accuracy. As shown in Fig.~\ref{fig3}, the feature map $F$ is firstly reshaped and permuted to $\mathbb{R}^{HWS\times C}$ and class centers are concatenated along the last dimension. The matrix multiplication is then executed between feature map $F\in \mathbb{R}^{HWS\times C}$ and class center $F^{class}\in \mathbb{R}^{C\times 1}$. The results are normalized to generate affinity map $A$ as follows:
\begin{equation}\label{e4}
A_{j}^{class} = \frac{exp(F^{class}\cdot F(j))}{\sum_{i}^{classes}exp(F^{i}\cdot F(j))}
\end{equation}
where $A_{j}^{class}$ denotes the similarity between the feature vectors in position $j$ and clustering center of $class$. The class set $classes=\{foreground,background\}$, which also can be extended to multi-class. Affinity map is then multiplied by the transposed feature vectors of class centers to obtain the attention feature map that is further element-wisely added to feature map $F$. The generation of novel feature map is formulated as follows:
\begin{equation}\label{e5}
F_{j}^{atten} = F(j)+\sum_{i}^{classes}A_{j}^{i}\cdot F^{i}
\end{equation}

As presented in formulas \ref{e4} and \ref{e5}, two points with similar features are more likely to be assigned the same labels, which achieves the continuous segmentation results of health cartilage while adjacent points with dissimilar features are hardly to be classified into the same class that ensures the discontinuity of defective cartilage. PCAM is a flexible plug-in module, the output of which will be the input of next layer in segmentation network as shown in Fig.~\ref{fig2}. Comparing to the self-attention module~\cite{ref_article4}, the float point operations of PCAM is reduced to $(2C-1)\times N\times HWS$ in a $C\times H\times W\times S$ feature map, where $N$ is the number of class. In PCAM, the side-output result indicates the class distribution and determines the accuracy of class center estimation. For ensuring the consistency between side-output result and true mask, the auxiliary deep supervision is adopted.
To relieve the class-imbalanced problem that knee cartilage occupies a much smaller area compared with background, the Dice loss and Cross-entropy loss are employed for supervision. The total loss is described as follows:
\begin{equation}\label{e6}
Loss_{Total} = Loss_{Dice} + Loss_{Cross-entropy}
\end{equation}
\section{Experiment}
\subsubsection{Materials and Evaluation Metrics}
The proposed method is validated on the OAI-ZIB Dataset\footnote[1]{\url{https://nda.nih.gov/oai/}}. This public dataset includes 507 3D DESS MR data with 81120 slices. The pixel spacing is 0.3645mm $\times$ 0.3645mm and the slice thickness is 0.7mm for all volumetric data. For each volumetric data, it contains 160 slices in 384 $\times$ 384. The 2-fold cross validation approach is applied in experiment for evaluating the performance of methods. To verify the effectiveness of PCAM, the ablation study is executed as well. Dice similarity coefficient (DSC), average symmetric surface distance (ASSD) and volumetric overlap error (VOE) are adopted for the comparison between the predicted results and ground truth. 
\subsubsection{Implementation Details}
In training phase, the batch size is set to 4 and the initial learning rate is set to 0.01 by Adam optimizer with 0.95 decay rate for every epoch. The training phase would be stopped when the improvement of DSC on validation set is no more than 0.0001 in continuous 10 epochs. The structure element is a $3\times 3$ mask with all elements set to 1. To protect the structure of cartilage, the morphological operation is executed only once. For data augmentation, the elastic deformation, random rotation, random-/center-cropping and random clip contribute to spatial transformation. Moreover, the gamma transformation, Gaussian noise and contrast adjustment are applied to enrich the gray distribution of training data. For mini-batch training in segmentation network, the batch normalization and ReLU are substituted by instance normalization and Leaky ReLU, respectively. The networks are trained and tested on NVIDIA 2080Ti with 11 GB video memory.
\subsubsection{Experimental results}
As the cascaded model (e.g.~\cite{ref_article2}-~\cite{ref_proc2}, ~\cite{ref_article5}) is composed of single segmentation models, the segmentation performance of cascaded model depends on its sub-networks with complex computational process and huge computation burdens. In this experiment, several classical segmentation networks with different schemes are evaluated. First of all, the baseline model is derived from V-Net~\cite{ref_proc4} without any modification. The second segmentation model is devised with the joint learning by the generative adversarial network (GAN) whose architecture and training process are adjusted on the basis of~\cite{ref_proc2}. Because the memory usage of attention module in ~\cite{ref_proc6} exceeds the limitation of GPU, the third model, that is the combination of baseline model and the attention module~\cite{ref_proc6}, is re-designed with the help of side-output and auxiliary supervision. The fourth model is the baseline with proposed PCAM plugged between the third and the fourth upsampling layer as Fig.~\ref{fig2}. The fifth model is the primary nnU-Net~\cite{ref_article5}, which obtained the best segmentation results on several medical image segmentation challenges. In the last model, the nnU-Net is combined with the proposed PCAM in the same location as the fourth model.

\begin{table}[h]
\caption{Quantitative comparisons among segmentation methods with evaluation metrics (DSC, VOE and ASSD) by mean and std values.}\label{tab1}
\centering
\begin{tabular}{|c|ccc|ccc|}
\hline
\multirow{2}{*}{Model} & \multicolumn{3}{c|}{Femoral Cartilage} & \multicolumn{3}{c|}{Tibial Cartilage} \\ \cline{2-7} 
 & \multicolumn{1}{c|}{DSC(\%)} & \multicolumn{1}{c|}{VOE(\%)} & ASSD(mm) & \multicolumn{1}{c|}{DSC(\%)} & \multicolumn{1}{c|}{VOE(\%)} & ASSD(mm)\\ \hline
Baseline~\cite{ref_proc4}& \multicolumn{1}{c|}{\begin{tabular}[c]{@{}c@{}}87.71\\ 2.77\end{tabular}} & \multicolumn{1}{c|}{\begin{tabular}[c]{@{}c@{}}21.79\\ 4.28\end{tabular}} & \begin{tabular}[c]{@{}c@{}}0.2259\\ 0.1020\end{tabular} & \multicolumn{1}{c|}{\begin{tabular}[c]{@{}c@{}}84.11\\ 3.93\end{tabular}} & \multicolumn{1}{c|}{\begin{tabular}[c]{@{}c@{}}27.23\\ 5.79\end{tabular}} & \begin{tabular}[c]{@{}c@{}}0.2287\\ 0.1067\end{tabular} \\ \hline
Tan et al.~\cite{ref_proc2}& \multicolumn{1}{c|}{\begin{tabular}[c]{@{}c@{}}87.86\\ 2.84\end{tabular}} & \multicolumn{1}{c|}{\begin{tabular}[c]{@{}c@{}}21.53\\ 4.40\end{tabular}} & \begin{tabular}[c]{@{}c@{}}0.2390\\ 0.1221\end{tabular} & \multicolumn{1}{c|}{\begin{tabular}[c]{@{}c@{}}84.08\\ 4.04\end{tabular}} & \multicolumn{1}{c|}{\begin{tabular}[c]{@{}c@{}}27.25\\ 5.94\end{tabular}} & \begin{tabular}[c]{@{}c@{}}0.2643\\ 0.1402\end{tabular} \\ \hline
Zhang et al.~\cite{ref_proc6}& \multicolumn{1}{c|}{\begin{tabular}[c]{@{}c@{}}87.91\\ 2.82\end{tabular}} & \multicolumn{1}{c|}{\begin{tabular}[c]{@{}c@{}}21.45\\ 4.38\end{tabular}} & \begin{tabular}[c]{@{}c@{}}0.2235\\ 0.1121\end{tabular} & \multicolumn{1}{c|}{\begin{tabular}[c]{@{}c@{}}84.57\\ 4.16\end{tabular}} & \multicolumn{1}{c|}{\begin{tabular}[c]{@{}c@{}}26.52\\ 6.20\end{tabular}} & \begin{tabular}[c]{@{}c@{}}0.2374\\ 0.0907\end{tabular} \\ \hline
Baseline~\cite{ref_proc4}+PCAM & \multicolumn{1}{c|}{\begin{tabular}[c]{@{}c@{}}88.45\\ 2.57\end{tabular}} & \multicolumn{1}{c|}{\begin{tabular}[c]{@{}c@{}}20.62\\ 4.04\end{tabular}} & \begin{tabular}[c]{@{}c@{}}\textbf{0.2058}\\ 0.0900\end{tabular} & \multicolumn{1}{c|}{\begin{tabular}[c]{@{}c@{}}85.15\\ 4.19\end{tabular}} & \multicolumn{1}{c|}{\begin{tabular}[c]{@{}c@{}}25.61\\ 6.31\end{tabular}} & \begin{tabular}[c]{@{}c@{}}0.2172\\ 0.1072\end{tabular} \\ \hline
nnU-Net~\cite{ref_article5}& \multicolumn{1}{c|}{\begin{tabular}[c]{@{}c@{}}89.03\\ 2.73\end{tabular}} & \multicolumn{1}{c|}{\begin{tabular}[c]{@{}c@{}}19.21\\ 4.37\end{tabular}} & \begin{tabular}[c]{@{}c@{}}0.2551\\ 0.3206\end{tabular} & \multicolumn{1}{c|}{\begin{tabular}[c]{@{}c@{}}86.0\\ 4.52\end{tabular}} & \multicolumn{1}{c|}{\begin{tabular}[c]{@{}c@{}}24.29\\ 6.73\end{tabular}} & \begin{tabular}[c]{@{}c@{}}\textbf{0.2117}\\ 0.1074\end{tabular} \\ \hline
nnU-Net~\cite{ref_article5}+PCAM & \multicolumn{1}{c|}{\begin{tabular}[c]{@{}c@{}}\textbf{89.35}\\ 2.69\end{tabular}} & \multicolumn{1}{c|}{\begin{tabular}[c]{@{}c@{}}\textbf{19.14}\\ 4.32\end{tabular}} & \begin{tabular}[c]{@{}c@{}}0.2389\\ 0.3196\end{tabular} & \multicolumn{1}{c|}{\begin{tabular}[c]{@{}c@{}}\textbf{86.14}\\ 4.43\end{tabular}} & \multicolumn{1}{c|}{\begin{tabular}[c]{@{}c@{}}\textbf{24.08}\\ 6.63\end{tabular}} & \begin{tabular}[c]{@{}c@{}}0.2165\\ 0.1242\end{tabular} \\ \hline
\end{tabular}
\end{table}

The experimental results are demonstrated on  Table.~\ref{tab1}. For adversarial learning scheme, the results are imporved on femoral cartilage segmentation but failed on the tibial cartilage. The structure characteristics that tibial cartilage occupies much fewer pixels may cause the adversarial learning scheme ineffective. For attention module in the third model, it captures the long-range contextual information that is more suitable for elongated cartilage segmentation. Furthermore, the segmentation performance of both baseline model and nnU-Net are improved on condition that the PCAM is plugged in the network which proves the effectiveness of the proposed self-attention module. Comparing to the primary baseline model, the combination of baseline model and PCAM can achieve the continuous and accurate segmentation, which is shown in Fig~\ref{fig1}. To quantify the continuity correctness of segmentation results, 0-dimension Betti number error is calculated slice-by-slice~\cite{ref_article6}. Under this metric, the average continuity errors of nnU-Net+PCAM model are $0.1323(\pm 0.43)$ in femoral cartilage and $0.1267(\pm 0.48)$ in tibial cartilage compared to nnU-Net with $0.1792(\pm 0.52)$ in femoral cartilage and $0.1358(\pm 0.49)$ in tibial cartilage. T-test was conducted between the methods with and without PCAM on segmentation continuity. We obtained p-value<0.01 on both femoral cartilage and tibial cartilage segmentation tasks, indicating that PCAM improves the segmentation continuity significantly. 

The decoder of the baseline model contains abundant semantic and spatial information, which is suitable to insert PCAM. However, there are three locations among four upsampling layers of decoder that the PCAM can be plugged in. To find out which upsampling layer of the decoder is optimized, the three locations are remarked as $1, 2, 3$ from low resolution to high resolution and plugged with PCAM, respectively. The experimental results are shown in Table.~\ref{tab2}. It can be seen that the PCAM plugged between the third and the fourth upsampling layer with the highest resolution has obtained the best segmentation results.
\begin{table}[h]
\caption{Ablation experiment on $PCAM$. $PCAMNet^{i}$ represents that the model is plugged with PCAM behind the $i th$ upsampling layer (PCAMNet=Baseline+PCAM). The performance is evaluated on three metrics by mean and std values.}\label{tab2}
\centering
\begin{tabular}{|c|ccc|ccc|}
\hline
\multirow{2}{*}{} & \multicolumn{3}{c|}{Femoral Cartilage}                                                                                                                                                                              & \multicolumn{3}{c|}{Tibial Cartilage}                                                                                                                                                                               \\ \cline{2-7} 
                  & \multicolumn{1}{c|}{DSC(\%)}                                                     & \multicolumn{1}{c|}{VOE(\%)}                                                   & ASSD(mm)                                                    & \multicolumn{1}{c|}{DSC(\%)}                                                     & \multicolumn{1}{c|}{VOE(\%)}                                                   & ASSD(mm)                                                    \\ \hline
$PCAMNet^{1}$  & \multicolumn{1}{c|}{\begin{tabular}[c]{@{}c@{}}87.78\\ 2.87\end{tabular}} & \multicolumn{1}{c|}{\begin{tabular}[c]{@{}c@{}}21.67\\ 4.47\end{tabular}} & \begin{tabular}[c]{@{}c@{}}0.2555\\ 0.1853\end{tabular} & \multicolumn{1}{c|}{\begin{tabular}[c]{@{}c@{}}84.57\\ 4.16\end{tabular}} & \multicolumn{1}{c|}{\begin{tabular}[c]{@{}c@{}}26.51\\ 6.20\end{tabular}} & \begin{tabular}[c]{@{}c@{}}0.2374\\ 0.0907\end{tabular} \\ \hline
$PCAMNet^{2}$  & \multicolumn{1}{c|}{\begin{tabular}[c]{@{}c@{}}87.90\\ 2.71\end{tabular}}  & \multicolumn{1}{c|}{\begin{tabular}[c]{@{}c@{}}21.49\\ 4.20\end{tabular}} & \begin{tabular}[c]{@{}c@{}}0.2438\\ 0.1451\end{tabular} & \multicolumn{1}{c|}{\begin{tabular}[c]{@{}c@{}}84.65\\ 4.14\end{tabular}} & \multicolumn{1}{c|}{\begin{tabular}[c]{@{}c@{}}26.39\\ 6.15\end{tabular}} & \begin{tabular}[c]{@{}c@{}}0.2518\\ 0.1746\end{tabular} \\ \hline
$PCAMNet^{3}$ & \multicolumn{1}{c|}{\begin{tabular}[c]{@{}c@{}}\textbf{88.45}\\ 2.57\end{tabular}} & \multicolumn{1}{c|}{\begin{tabular}[c]{@{}c@{}}\textbf{20.62}\\ 4.04\end{tabular}} & \begin{tabular}[c]{@{}c@{}}\textbf{0.2058}\\ 0.0900\end{tabular}   & \multicolumn{1}{c|}{\begin{tabular}[c]{@{}c@{}}\textbf{85.15}\\ 4.19\end{tabular}} & \multicolumn{1}{c|}{\begin{tabular}[c]{@{}c@{}}\textbf{25.61}\\ 6.31\end{tabular}} & \begin{tabular}[c]{@{}c@{}}\textbf{0.2172}\\ 0.1072\end{tabular} \\ \hline
\end{tabular}
\end{table}

In PCAM, the side-output is used to calculate the center of each class. But the original side-output $S_{pred}$ contains many false positives on the area around the boundary of segmentated knee cartilage as shown in Fig.~\ref{fig1}. In order to filter out segmentation mistakes, the Erode operations (achieved by $maxpooling$) are applied in this research. The modified side-output $S_{pred}^{'}$ has higher precision results that are shown in Table.~\ref{tab3}.
\begin{table}[h]
\caption{Quantitative comparion by mean and std precision between original side-output $S_{pred}$ results and side-output results $S_{pred}^{'}$ modified by morphological operations. ($Precision = \frac{True Positives}{True Positives + False Positives}\times 100\%$)}\label{tab3}
\centering
\begin{tabular}{|c|cccc|}
\hline
\multirow{3}{*}{} & \multicolumn{4}{c|}{Precision(\%)}                                                                                                                                                                                                                                                                       \\ \cline{2-5} 
                  & \multicolumn{2}{c|}{Femoral Cartilage}                                                                                                                      & \multicolumn{2}{c|}{Tibial Cartilage}                                                                                                  \\ \cline{2-5} 
                  & \multicolumn{1}{c|}{Foreground}                                              & \multicolumn{1}{c|}{Background}                                              & \multicolumn{1}{c|}{Foreground}                                              & Background                                              \\ \hline
$S_{pred}$             & \multicolumn{1}{c|}{\begin{tabular}[c]{@{}c@{}}87.56\\ 4.18\end{tabular}} & \multicolumn{1}{c|}{\begin{tabular}[c]{@{}c@{}}99.83\\ 0.05\end{tabular}} & \multicolumn{1}{c|}{\begin{tabular}[c]{@{}c@{}}83.32\\ 7.51\end{tabular}} & \begin{tabular}[c]{@{}c@{}}99.95\\ 0.02\end{tabular} \\ \hline
$S_{pred}^{'}$            & \multicolumn{1}{c|}{\begin{tabular}[c]{@{}c@{}}\textbf{98.13}\\ 1.75\end{tabular}} & \multicolumn{1}{c|}{\begin{tabular}[c]{@{}c@{}}\textbf{99.98}\\ 0.02\end{tabular}} & \multicolumn{1}{c|}{\begin{tabular}[c]{@{}c@{}}\textbf{94.77}\\ 5.98\end{tabular}} & \begin{tabular}[c]{@{}c@{}}\textbf{99.99}\\ 0.01\end{tabular} \\ \hline
\end{tabular}
\end{table}
\section{Conclusion}
In this research, we proposed a novel self-attention module PCAM for ensuring accurate continuous knee cartilage segmentation. The proposed PCAM captures long-range contextual information to relieve limited receptive field brought by convolution filters in neural networks. Besides, the proposed PCAM brings less computational burdens and is flexible to be plugged in any encoder-decoder structured segmentation neural networks. The experimental results show that the proposed method achieves an accurate segmentation both on femoral and tibial cartilage in 3D MR data and has a potential application in future.
\subsubsection{Acknowledgements} This work was supported by the National Natural Science Foundation of China under Grant 62001144 and Grant 62001141, and by Science and Technology Innovation Committee of Shenzhen Municipality under Grant JCYJ20210324131800002 and RCBS20210609103820029.

\end{document}